\documentclass[aip,graphicx,reprint]{revtex4-1}

\usepackage{graphicx}
\usepackage{dcolumn}% Align table columns on decimal point
\usepackage{bm}% bold math
\usepackage{amsmath}
\usepackage{amssymb}
\usepackage{siunitx}
\usepackage[T1]{fontenc}
\usepackage{mathptmx}
\usepackage{etoolbox}
\usepackage{99_IMS_constants}

\DeclareSIUnit{\belmilliwatt}{Bm}
\DeclareSIUnit{\dBm}{\deci\belmilliwatt}
\DeclareSIUnit{\pH}{\pico\henry}

\makeatletter
\def\@email#1#2{%
 \endgroup
 \patchcmd{\titleblock@produce}
  {\frontmatter@RRAPformat}
  {\frontmatter@RRAPformat{\produce@RRAP{*#1\href{mailto:#2}{#2}}}\frontmatter@RRAPformat}
  {}{}
}
\makeatother
\begin{document}

\title{SQUID-based superconducting microcalorimeter with in-situ tunable gain}

\author{C. Schuster}%
 \email{constantin.schuster@kit.edu}
\affiliation{Institute of Micro- and Nanoelectronic Systems, Karlsruhe Institute of Technology, Hertzstrasse 16, Building 06.41, D-76187 Karlsruhe, Germany}

\author{S. Kempf}%
\affiliation{Institute of Micro- and Nanoelectronic Systems, Karlsruhe Institute of Technology, Hertzstrasse 16, Building 06.41, D-76187 Karlsruhe, Germany}
\affiliation{Institute for Data Processing and Electronics, Karlsruhe Institute of Technology, Hermann-von-Helmholtz-Platz 1, Building 242, D-76344 Eggenstein-Leopoldshafen}

\date{\today}

\begin{abstract}
Cryogenic microcalorimeters are outstanding tools for X-ray spectroscopy due to their unique combination of excellent energy resolution and close to $100\,\%$ detection efficiency. While well-established microcalorimeter concepts have already proven impressive performance, their energy resolution has yet to improve to be competitive with cutting-edge wavelength-dispersive grating or crystal spectrometers. We hence present an innovative SQUID-based superconducting microcalorimeter with an in-situ tunable gain as alternative concept that is based on the strong temperature dependence of the magnetic penetration depth of a superconductor operated close to its critical temperature. Measurements using a prototype device show no sign for any hysteresis effects that often spoil the performance of superconducting microcalorimeters. Moreover, our predictions of the achievable energy resolution show that a competitive energy resolution $\mathcal{O}(300\,\mathrm{meV})$ with a suitable combination of absorber and sensor material should be easily possible.
\end{abstract}

\maketitle

Cryogenic microcalorimeters such as superconducting transition-edge sensors (TESs) \cite{Irwin2005,Ullom2015} or metallic magnetic calorimeters (MMCs) \cite{Fleischmann2005,Kempf2018} have proven to be outstanding devices to measure the energy of X-ray photons with utmost precision. They rely on sensing the change in temperature of an X-ray absorber upon photon absorption using an extremely sensitive thermometer that is based on either a superconducting (TES) or a paramagnetic (MMC) sensor material. Thanks to their exceptional combination of an outstanding energy resolution and a quantum efficiency close to $100\,\%$, they are highly attractive for X-ray emission spectroscopy (XES), both at synchrotron light sources and in a laboratory environment \cite{Friedrich2006, Doriese2016}. They strongly relax the requirements on X-ray beam intensity as compared to state-of-the-art wavelength-dispersive X-ray spectrometers based on diffraction gratings or bent crystals \cite{Uhlig2015}. Moreover, they cover the entire tender X-ray range \cite{Uhlig2015} that is hardly accessible with both, grating and crystal spectrometers, and allow studying strongly diluted as well as radiation-sensitive samples that can, even at the most brilliant synchrotron light sources, only be investigated with greatest efforts \cite{Friedrich2006, Doriese2016}.

The best TES- and MMC-based detectors to date achieve an energy resolution $\dEFWHM$ of $0.72\,\mathrm{eV}$ for $1.5\,\mathrm{keV}$ photons \cite{Lee2015} and of $1.25\,\mathrm{eV}$ for $5.9\,\mathrm{keV}$ photons \cite{Krantz2023} and provide a quantum efficiency close to $100\,\%$ in this energy range. In terms of resolution, they come close to the resolution of wavelength-dispersive X-ray spectrometers while offering an orders of magnitude higher efficiency. But despite of this great success, both detector types face some challenges that presently prevents these detectors to reach an energy resolution in the range of $100\,\mathrm{meV}$ as required for investigating vibrations or $d$-$d$-excitations in soft X-ray spectroscopy or resonant inelastic X-ray scattering. It might hence be worthwhile to look for alternate detector concepts in addition to advancing the present best technologies.

\begin{figure}
  \includegraphics[width=0.95\columnwidth]{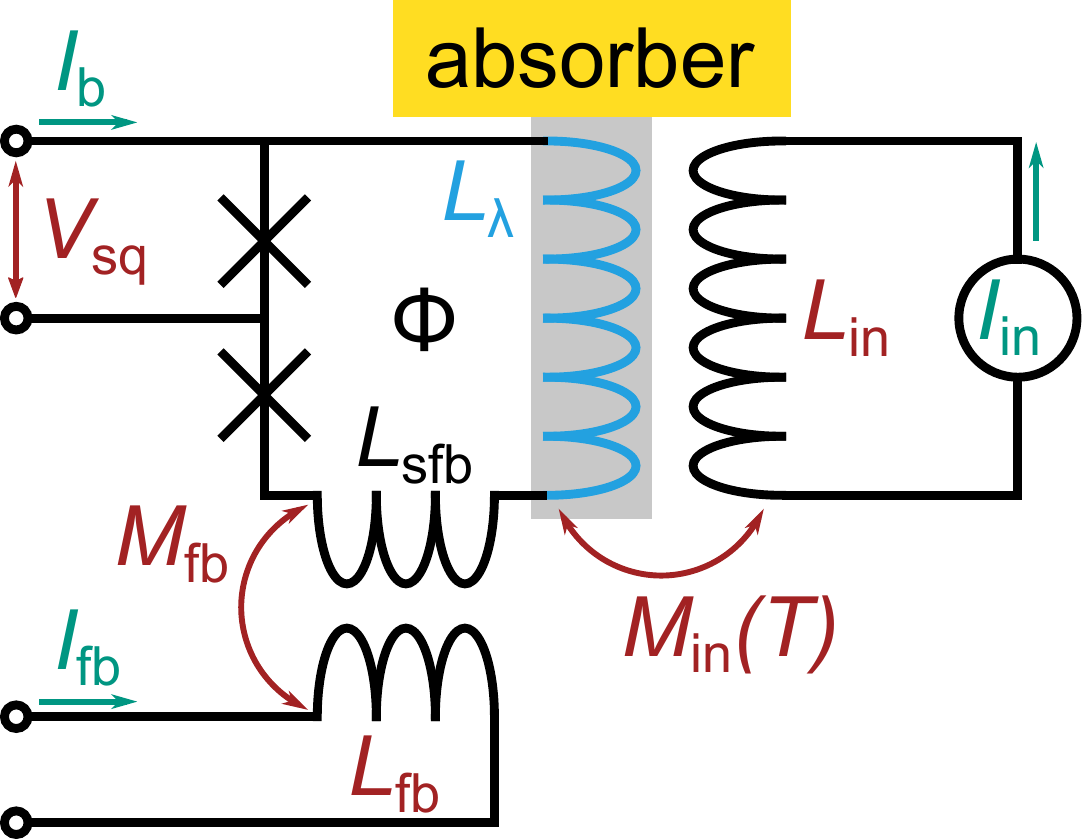}
  \caption{Schematic circuit diagram of a {\lSQ}. Inductors depicted in black are made from a superconducting material with critical temperature $\Tc$. The $\lambda$-coil displayed in blue is made from a different superconducting material with critical temperature $\Tl \ll \Tc$ and is in strong thermal contact with the absorber. The device is operated at temperature $T_0 \lessapprox \Tl$. A current source runs a constant current $I_\mathrm{in}$ through an input coil with inductance $\Lin$. The $\lambda$-coil and the input coil are inductively coupled via the mutual inductance $\Min(T)$ that is temperature-dependent because of the temperature dependence of the magnetic penetration depth $\lambda(T)$ of the $\lambda$-coil.}
  \label{fig:LSQ_shematic}
\end{figure}

\begin{figure*}
  \includegraphics[width=0.7\textwidth]{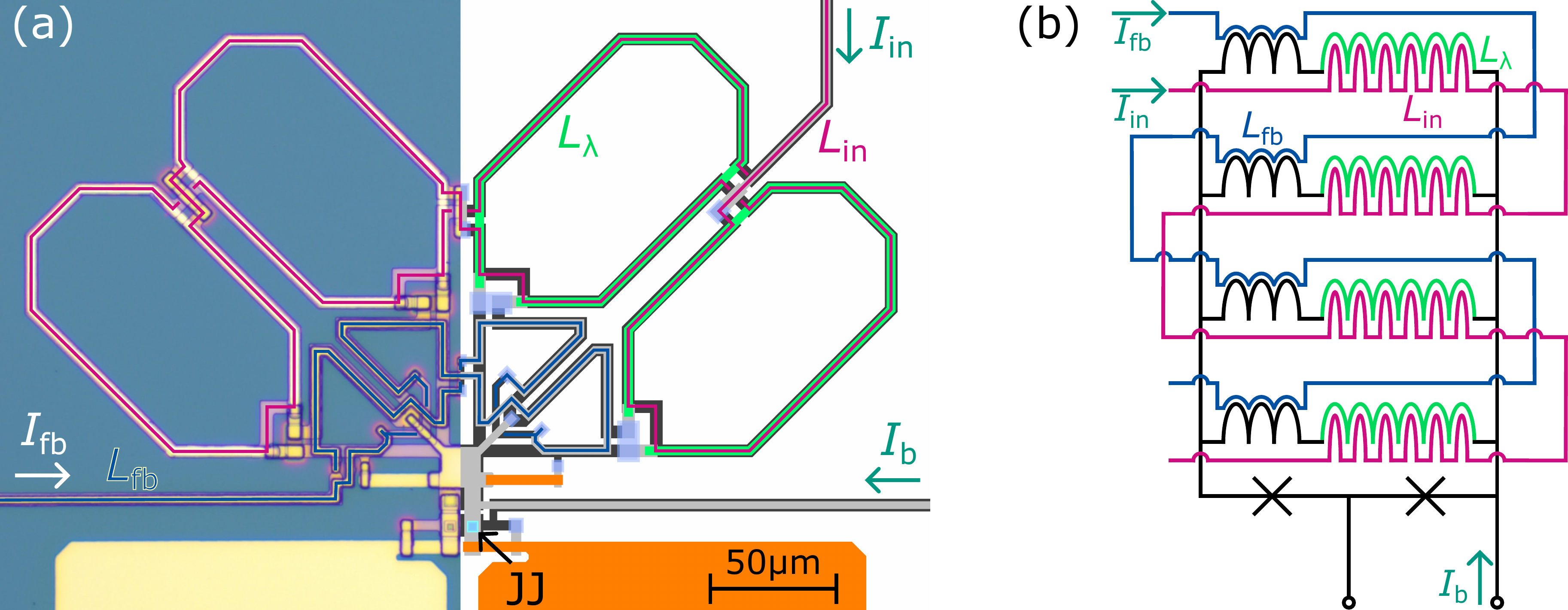}
  \caption{(a) Micrograph (left) and design layout (right) as well as (b) equivalent circuit diagram of our prototype $\lambda$-SQUID. The $\lambda$-coil (green in the design and circuit diagram, bright yellow in the micrograph) is made from Al ($\Tl \simeq \SI{1.2}{\K}$), while the other parts of the device are made from Nb ($\Tc \simeq \SI{8.9}{\K}$). In (a), the paths of the input and flux biasing coils are traced in magenta and blue, respectively.}
  \label{fig:design}
\end{figure*}

Here, we present an innovative SQUID-based microcalorimeter with in-situ tunable gain that appears to be particularly suited for high-resolution soft X-ray spectroscopy. It is based on the strong temperature dependence of the magnetic penetration depth $\lambda(T)$ of a superconducting material that is operated close to its critical temperature $\Tc$. Fig.~\ref{fig:LSQ_shematic} depicts a simplified schematic circuit diagram of our microcalorimeter which we denote as {\lSQ}. It largely resembles a conventional low-$T_\mathrm{c}$ dc-SQUID that comprises two identical resistively shunted Josephson tunnel junctions with critical current $\Ic$ and normal state resistance $R$ as well as a closed superconducting loop formed by two inductors with inductances $\Ll$ and $\Lsfb$, respectively. The Josephson junctions, the inductor $\Lsfb$ as well as the superconducting wiring are made from a superconducting material with critical temperature $\Tc \gg T_0$ that is much greater than the operating temperature $T_0$ of the microcalorimeter. In contrast, the inductor $\Ll$ which we will refer to as $\lambda$-coil is  made from a superconducting material with much lower critical temperature $\Tl \gtrapprox T_0$ that exceeds only barely the device operating temperature $T_0$. It is worth mentioning that for a practical device the operation temperature will be chosen according to the transition temperature of the $\lambda$-coil, and not vice versa. The $\lambda$-coil is coupled to the input coil with inductance $\Lin$ via the mutual inductance $\Min = k\sqrt{\Ll\Lin}$. Here, $k$ denotes the magnetic coupling factor. Since the $\lambda$-coil is operated near its critical temperature, its magnetic penetration depth $\lambda(T)$ depicts a strong temperature dependence, resulting in a temperature dependence of its inductance $\Ll(T) = \Llg(T) + \Llk(T)$. Here, $\Llg(T)$ and $\Llk(T)$ represent the geometric and kinetic contributions to the total coil inductance. The magnetic penetration depth influences the geometric inductance $\Llg(T)$ via the distribution of current density in the cross-section of the inductor, which ultimately causes a temperature dependence of the mutual inductance $\Min(T)$. 

The device is biased either with a constant current $\Ib$, with the resulting voltage drop $\Vsq$ across the device being used as output signal, or with a constant voltage $V_\mathrm{b}$, the current $I_\mathrm{SQ}$ running through the devices acting as output signal. In any case, the output signal depends on the total magnetic flux $\Phi_\mathrm{tot}$ threading the SQUID loop. A current source supplying a constant current $\Iin$ is connected to the input coil. This current induces a magnetic flux $\Phi(T) = \Min(T) \Iin$ within the SQUID loop that is temperature dependent via the temperature dependence of the mutual inductance $\Min(T)$. 

By attaching a suitable X-ray absorber to the $\lambda$-coil, the absorption of an X-ray photon causes a temperature rise of this coil that is transduced into a change of magnetic flux within the SQUID loop. The latter can be sensed as a change of output signal. An additional coil with inductance $\Lfb$ is inductively coupled to the {\lSQ} via the inductor $\Lsfb$ and mutual inductance $\Mfb$. This coil allows for additional control of the magnetic flux threading the SQUID loop by running an externally controlled current $\Ifb$ through it.

\begin{figure*}
  \includegraphics[width=0.8\textwidth]{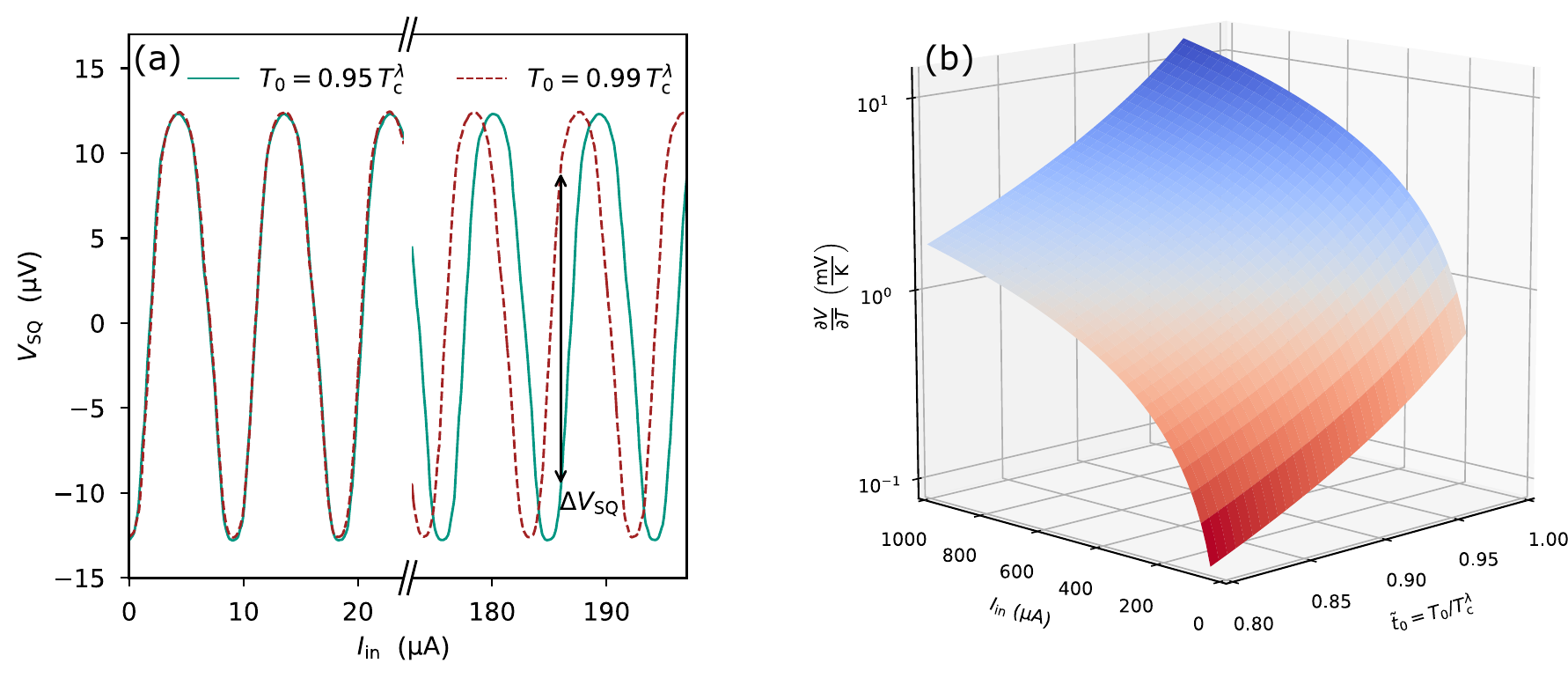}
  \caption{(a) Output voltage $\Vsq$ of our test device versus the input current $\Iin$ for two different operation temperatures. The change in $\Min$ is small, but with a sufficiently large choice of the input current $\Iin$ a sizeable signal $\Delta \Vsq$ can be achieved. (b) Gain coefficient $\partial \Vsq/\partial T$ versus the input current $\Iin$ and the reduced operation temperature $\tilde{t}_0 = T_0/\Tl$.}
  \label{fig:LSQ_VPHI}
\end{figure*}

For the following discussion, we assume a constant current bias. In this case, the relationship between output signal voltage $\Vsq$ and temperature $T$ is defined by the gain coefficient

\begin{equation}
    \frac{\partial \Vsq}{\partial T} =  \frac{\partial \Vsq}{\partial \Phi} \frac{\partial \Min}{\partial T} \Iin,
    \label{gainfactor}
\end{equation}
where the flux-to-voltage transfer coefficient can be estimated by $\partial \Vsq/\partial \Phi = R/\Lsq$ for an optimized dc-SQUID \cite{Tesche1977}. Here, $\Lsq$ denotes the total SQUID inductance. From Eq.~\ref{gainfactor}, the tunability of the gain factor gets clearly apparent: With $\Iin$ being limited solely by the ampacity of the input coil and the temperature dependence $\partial \Min/\partial T$ depending on the operation temperature, the gain coefficient can be varied by both, the input current and operation temperature, over a wide range. Moreover, with a sufficient choice of the current $\Iin$, the gain coefficient can be made suitably large, even if $\frac{\partial \Min}{\partial T}$ itself is rather small. In addition, as the current $\Iin$ can be changed fast, e.g. by external control electronics, the gain can be even tuned in-situ, i.e. even during the acquisition of an detection event. This enables new degrees of freedom during readout, e.g. by ensuring that the dynamic range of analogue to digital converters in the readout chain is always fully utilised.

We fabricated a prototype device that is depicted in Fig.~\ref{fig:design}a). It is based on the design of one of our custom-made dc-SQUIDs for MMC readout and features four superconducting loops that are connected in parallel (Fig.~\ref{fig:design}b). Each loop is inductively coupled to a flux biasing as well as an input coil where the coil arrangement is engineered such that the mutual inductance between these coils is greatly reduced. The majority of the device including the actual input coil with inductance $\Lin$ and flux biasing coil with inductance $\Lfb$ is fabricated from niobium having a critical temperature $\Tc \simeq 8.9\,\mathrm{K}$. The part of the SQUID loop that is coupled to the input coil acts as $\lambda$-coil and is made from aluminium with critical temperature $\Tl \simeq 1.2\,\mathrm{K}$. We note that aluminium is an unfavourable sensor material for a real microcalorimeter (as the thermal noise at $\simeq 1.2\,\mathrm{K}$ would prevent to reach an energy resolution in the range of $1\,\mathrm{eV}$ or below). However, this choice was triggered by our intention to use this device only for proving the suitability of our microcalorimeter concept and to estimate its sensitivity as well as the fact that an easy-to-deposit elemental superconductor with $T_\mathrm{c}$ in the range of $50\text{-}100\,\mathrm{mK}$ doesn't exist. We consequently postponed the development of a sophisticated deposition process for a superconducting material with suitable $T_\mathrm{c}$ and also omitted the particle absorber. A full working microcalorimeter comprising a low-$T_\mathrm{c}$ material for the $\lambda$-coil as well as a suitable X-ray absorber will hence be subject of future work.

We comprehensively characterized our prototype device in a $^3$He/$^4$He dilution refrigerator that was tweaked to smoothly run up to temperatures of about $1.5\,\mathrm{K}$. We used a resistive heater mounted at the mixing-chamber platform to control the operation temperature and biased and read out the detector using a direct-coupled high-speed dc-SQUID electronics \cite{Drung2006}. Fig.~\ref{fig:LSQ_VPHI}a) shows the measured output voltage $\Vsq$ across the device versus the applied input current $\Iin$ at two different operating temperatures $T_0$ close to the critical temperature $\Tl$ of the $\lambda$-coil. The periodic response of dc-SQUIDs is clearly visible. For small input currents, the tiny temperature-induced change of the mutual inductance hardly creates a measured voltage change $\Delta \Vsq$. However, the larger $\Iin$, the larger $\Delta \Vsq$. This directly resembles the prediction according to Eq.~\ref{gainfactor} and proves the in-situ tunability of the gain factor. In Fig.~\ref{fig:LSQ_VPHI}b), the gain coefficient $\partial \Vsq/\partial T$ is depicted versus both the input current $\Iin$ and the reduced operation temperature $\tilde{t}_0 = T_0 / \Tl$, illustrating the large tuning range available with these two parameters.

To determine the temperature dependence of the mutual inductance $\Min(T)$ as well as the sensitivity coefficient $\partial \Vsq/\partial T$, we performed two different types of measurements: For \textit{static measurements}, we injected a linear current ramp into the input coil and monitored the output voltage $\Vsq$. The current change $\Delta \Iin$ for a flux change of a single flux quantum $\Phi_0$ then allows to determine the mutual inductance via $\Min = \Phi_0 / \Delta \Iin$. For \textit{flux ramp modulated\cite{Mates2012, Richter2021} (FRM) measurements}, we injected a constant input current $\Iin$ into the input coil and applied a periodic sawtooth-like current ramp with amplitude $I_\mathrm{ramp}$ and repetition rate $\framp$ to the flux biasing coil. We chose the ramp amplitude $I_\mathrm{ramp}$ such that in each cycle an integer multiple of flux quanta was induced into the {\lSQ}. In this mode of operation, the output voltage $\Vsq$ is modulated with a well-defined frequency. Any flux signal $\Phi$ that is quasi-static with respect to the ramp repetition rate $\framp$ is then transduced into a phase shift $\Theta = 2 \pi \Phi / \Phi_0$ of the periodic output voltage $\Vsq$. The flux contribution $\Phi(T) = \Min(T) \Iin$ resulting from the constant current within the input coil changes with temperature and can be derived from the phase shift $\Theta(T)$ of the {\lSQ} response. The latter can be determined by demodulation of the SQUID output voltage. It is worth mentioning that a change of the SQUID output voltage can be caused by a shift of the working point, too, and may affect the shape and offset of the periodic output voltage. However, the FRM phase and thus demodulated flux signal remain unaffected. The FRM method hence allows to distinguish different contributions affecting the SQUID voltage. 

\begin{figure*}
  \includegraphics[width=0.8\textwidth]{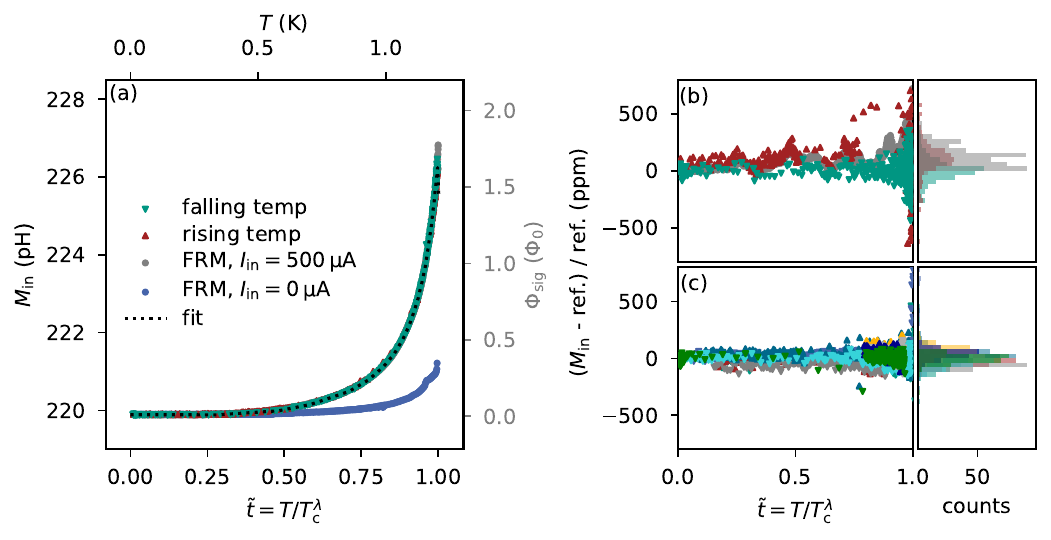}
  \caption{(a) Mutual inductance $\Min$ (left axis) as a function of reduced temperature $\tilde{t} = T/\Tl$ for two static measurements (performed with increasing and decreasing temperature, respectively) as well as an FRM measurement with $\Iin = \SI{500}{\mu A}$. Moreover, an empirical fit to these curves is shown as explained in the main text. Additionally, the signal flux $\Phi$ (right axis) is displayed for the FRM measurements. (b) Deviation of the curves for rising temperature (red, $\blacktriangledown$), falling temperature (green, $\blacktriangle$) and FRM with $\Iin = \SI{500}{\mu A}$ to a common, smoothed reference curve (ref.) to substantiate the excellent agreement. A histogram of the deviation data is provided also. The mean deviation from 0 is below $\SI{250}{ppm}$, indicating that no significant hysteresis is present between measuring during warm-up and cool-down. (c) is equivalent to (b), but for a second test device of identical design. For this device, six curves of $\Min(T)$ while cooling down ($\blacktriangledown$, various colours) and four curves while warming up ($\blacktriangle$, various colours) were acquired using the static method. Again, no statistically significant deviation is present.}
  \label{fig:Min}
\end{figure*}

Fig.~\ref{fig:Min}a) shows the measured mutual inductance $\Min$ of our prototype device as a function of reduced temperature $\tilde{t} = T/\Tl$. We have performed several static measurements while both, warming up and cooling down, in order to check for potential hysteresis effects that would potentially spoil the detector performance \cite{Stevenson2013}. The FRM measurement was performed with an injected current $\Iin = \SI{500}{\mu A}$ and served to affirm the results of the static measurements. The FRM curve can hardly be seen in Fig.~\ref{fig:Min}a) as it fully overlaps with the data from the static measurements, indicating excellent agreement. We performed an empirical fit of the measured curves using the function 
\begin{equation}
\begin{aligned}
f\left(t\right) &= \Min^0 \{ 1 - a \left[ 1 - z\left(\tilde{t}\right)\right]\} \\
z\left(t\right) &= \left(1 - \tilde{t}^4 \right)^{-1} \\
\end{aligned}
\end{equation}
The fit parameter $\Min^0$ describes the mutual inductance at zero temperature, and the constant $a$ quantifies the magnitude of the temperature change of $\Min(T)$. This fit is used for the prediction of noise performance below. The overall shape of the curve with the very strong increase in $\Min(T)$ close to $\Tl$ arises from the temperature dependence of $\lambda(T)$, which diverges at the critical temperature. To illustrate the excellent agreement between the three data sets and hence the absence of any hysteresis effects, their deviation from a common, smoothed reference curve is displayed in figure \ref{fig:Min}b). The deviation of some data points in the curve for rising temperature (red, $\blacktriangle$) is due to a thermal instability of our cryostat in the temperature range of $\SI{800}{mK}$ to $\SI{1}{K}$, leading to sudden and fast changes in temperature causing an offset between measured and actual temperature of the test device. Over a large temperature range, the deviation between data sets is lower than $\SI{250}{ppm}$, as can be seen from the histrogramm representation. Similarly, a total of ten static measurements were performed on a second, identical test device. Six of these were measured while cooling down, four curves while warming up. Again, their deviations to a smoothed reference curve were determined and are displayed in \ref{fig:Min}c). The related histogram proves again that no statistically relevant offset between warming up and cooling down is present and hence that no hysteresis effects occur.

We additionally performed an FRM measurement without applying an input current to check for parasitic inductance effects (see Fig.~\ref{fig:Min}a)). Since $\Iin=0$, we can not use the equation $\Phi(T) = \Min(T) \Iin$ to relate the signal flux $\Phi$ to a change in mutual inductance $\Min$. In this scenario, no flux signal is induced via the input coil. Instead, the measured signal is caused by the temperature dependence of the inductance $\Ll$ coupling to the bias current of the SQUID. The parasitic inductance effects are negligible as compared to the change of mutual inductance.

We used the acquired data on $\Min(T)$ to estimate the achievable energy resolution $\dEFWHM$ of a fully equipped microcalorimeter with $\Tl$ below $100\,\mathrm{mK}$. For this, we assume that the temperature dependence $\Min(\tilde{t})$ solely depends on the reduced temperature $\tilde{t} = T/\Tl$, i.e.
\begin{equation}
    \frac{dM}{dT} = \frac{\partial M}{\partial \tilde{t}} \frac{d \tilde{t}}{dT}.
\end{equation}
This assumption is well justified as the magnetic penetration depth of a superconducting material shows this scaling behaviour. We hence can use our measured data to extrapolate the energy resolution for a device comprising a superconducting material with lower critical temperature $\Tl$ than Al. If the noise of the readout chain reading out the detector is sufficiently low, for example by using a voltage bias of the detector as well as an $N$-dc SQUID series array as a first amplifier stage, there are two noise contributions to consider:  Thermal noise $\SEtd$ resulting from random energy fluctuations among the absorber, sensor and heat bath as well as SQUID noise $\SEsq$. The latter is given by
\begin{equation}
    \SEsq = \Svv \left( \frac{\partial \Vsq}{\partial T} \frac{\partial T}{\partial E} \right)^{-2}
\end{equation}
with the voltage noise $\Svv$ of the SQUID, the gain coefficient ${\partial \Vsq}/{\partial T}$ and the inverse total heat capacity $1/\Cdet = {\partial T}/{\partial E}$. For an optimized dc-SQUID\cite{Tesche1977}, the SQUID voltage noise can be readily estimated by $\Svv = 18 \kB T R$. To determine the gain coefficient
\begin{equation}
    \frac{\partial \Vsq}{\partial T} =  \frac{\partial \Vsq}{\partial \Phi} \frac{\partial \Min}{\partial \tilde{t}} \frac{\Iin}{\Tl},
\end{equation}
we deduce $\partial \Min/\partial \tilde{t}$ from the empirical fit in Fig.~\ref{fig:Min}a) by evaluating its derivative at the operating temperature $T_0$. The thermal noise $\SEtd$ among absorber, sensor and heat bath follows the relation \cite{McCammon2005, Fleischmann2005}
\begin{equation}
    \SEtd = \kB \Csen T^2 \left[ \frac{4  \left(1 - \beta \right) \tau_0}{1 + \left( 2 \pi \tau_0 f \right)^2} + \frac{4 \beta \tau_1}{1 + \left(2 \pi \tau_1 f\right)^2} \right].
\end{equation}
Here, $\tau_0$ and $\tau_1$ represent the signal rise and decay time, $\Csen$ is the heat capacity of the temperature sensor, and $\beta = \Csen/\Ctot$.
The energy resolution $\dEFWHM$ is then determined by the integral \cite{McCammon2005, Fleischmann2005}:
\begin{equation} \label{eq:energyres}
\dEFWHM = 2 \sqrt{2 \mathrm{ln}2} \left[ \int\limits_0^\infty  \frac{\left| p(f) \right|^2 }{\SEtd(f) + \SEsq(f)} \mathrm{d}f \right]^{-1/2}
\end{equation}
with the detector responsivity
\begin{equation}
\left| p(f) \right| = \frac{2 \beta \tau_1}{\sqrt{1 + \left( 2 \pi \tau_0 f \right)^2} \sqrt{1 + \left( 2 \pi \tau_1 f \right)^2}}
\end{equation}
For the achievable energy resolution, both noise contributions appear in the denominator of Eq.~\ref{eq:energyres}. To calculate the energy resolution caused by only one of these noise contributions as used in the comparison below, only the considered noise appears there.

\begin{figure}
  \includegraphics[width=1.0\columnwidth]{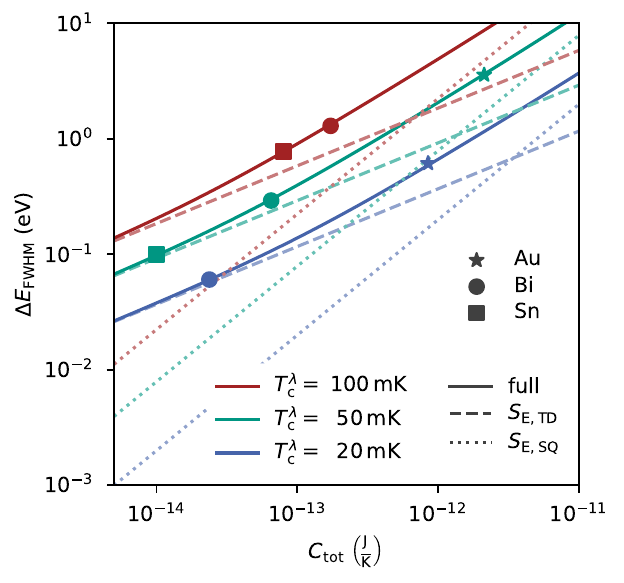}
  \caption{Predicted energy resolution $\dEFWHM$ versus total heat capacity $\Ctot$ of the detector. The achievable energy resolution (solid line) is displayed as well as the effective energy resolutions if we only consider the thermodynamic noise $\SEtd$ (dashed) or only the noise $\SEsq$ of the {\lSQ} (dotted), respectively. Curves for three values of $\Tl=\SI{100}{mK}$ (red), $\Tl=\SI{50}{mK}$ (green), and $\Tl=\SI{20}{mK}$ (blue) at a constant input current of $\Iin = \SI{500}{\mu A}$. Stars, circles and squares mark the full energy resolution when using absorbers made from Au, Bi and Sn, respectively, at the respective temperatures.}
  \label{fig:noise}
\end{figure}

Fig.~\ref{fig:noise} shows the predicted energy resolution(s) versus the total heat capacity $\Ctot$ of the detector for three different critical temperatures of the $\lambda$-coil. For these calculations, we assumed a conservative input current of $\Iin = \SI{500}{\mu A}$, rise and decay times of $\tau_0 = \SI{1}{\mu s}$ and $\tau_1 = \SI{1}{m s}$ 
\footnote{The rise and decay times can be set via the design of the thermal links between absorber, sensor and thermal bath. As the energy resolution scales like $\dEFWHM \propto (\tau_0 / \tau_1)^{1/4}$, a reduction of the decay time comes at a cost of increased thermodynamic energy fluctuations \cite{Fleischmann2005}. The values used here are typical for MMCs, but can be tuned to suit other applications.}
and $\beta=1/2$ as well as an operating temperature of $T_0 = 0.95 \Tl$.
The contributions from SQUID noise (dotted) and thermal noise (dashed) scale differently with the heat capacity, leading to a crossover point below which the thermal noise is dominating. It is worth mentioning that the SQUID noise scales with total heat capacity as the heat capacity and hence the geometric size of the $\lambda$-coil must match the absorber heat capacity. As expected, a decrease in $\Tl$ (and thus $T_0$) leads to an overall improvement of the energy resolution as well as a shift of the crossover point towards higher heat capacities. To give a specific example, we want to assume an X-ray detector with a sensitive detection area of $\SI{250}{\mu m} \times \SI{250}{\mu m}$ and a thickness $d$ of $\SI{5}{\mu m}$, $\SI{8.6}{\mu m}$ and $\SI{50}{\mu m}$ for the absorber materials Au, Bi and Sn, respectively. Thus, the stopping power is roughly identical for all materials and exceeds $\SI{99.99}{\percent}$ at photon energies up to $\SI{1}{\kilo eV}$. We select gold, bismuth and tin as absorber material as these materials are often used in the cryogenic microcalorimeter community \cite{Horansky2008,Brown2008,Krantz2023}. Fig.~\ref{fig:noise} shows that for a proper combination of absorber material and critical temperature of the sensor material, an energy resolution as low as $300\,\mathrm{meV}$ should be feasible. It should be mentioned that a Bi absorber operated close to $\SI{50}{\milli \kelvin}$ with a specific heat of $\SI{6.5e-14}{\joule \per \kelvin}$ would experience a temperature increase of $\SI{2.46}{\milli \kelvin}$ (or just below $\SI{5}{\percent}$ of $\Tl$) upon absorption of a $\SI{1}{\kilo eV}$ photon. It is thus expected to display some nonlinear behaviour, which may potentially be predicted theoretically and compensated for. Moreover, it should be noted that increasing the input current $\Iin$ suppresses the SQUID noise contribution further (cf. Eq.~\ref{gainfactor}). This allows, for example, to compensate for a potential readout noise degradation when using SQUID-based multiplexing techniques \cite{Irwin2009}.

In conclusion, we have presented an innovative concept for a SQUID-based superconducting microcalorimeter with an in-situ tunable gain. It is based on the strong temperature dependence of the magnetic penetration depth of a superconductor close to its critical temperature $\Tc$ that affects the mutual inductance $\Min$ between the SQUID loop and an input coil that is biased with a constant current. The latter can be easily tuned in-situ. This allows, for example, for compensating a potential noise degradation when using cryogenic multiplexing techniques. We have successfully designed, fabricated and characterised a prototype device using aluminium as sensor material to study the temperature dependence of $\Min$. We find that there is no sign for any hysteresis effects that often spoil the performance of superconducting microcalorimeters. Using this data, we have made predictions of the achievable energy resolution. We found that the lower the total specific heat $\Ctot$ of the detector, the easier it is to suppress the {\lSQ} noise below the thermal noise floor. More specifically, we found that an energy resolution $\mathcal{O}(300\,\mathrm{meV})$ with a suitable combination of absorber and sensor material is possible.

\begin{acknowledgments}
We would like to thank A. Stassen for his support during device fabrication and greatly acknowledge fruitful discussions with G. Jülg. We acknowledge financial support by the KIT Center of Elementary Particle and Astroparticle Physics (KCETA). Furthermore, C. Schuster acknowledges financial support by the Karlsruhe School of Elementary Particle and Astroparticle Physics: Science and Technology (KSETA).
%%%%

\end{acknowledgments}

\section*{Author Declarations}
\subsection*{Conflict of Interest}
The authors have no conflicts to disclose.

\subsection*{Author Contributions}
%{\color{red}
{\bf Constantin Schuster:} Conceptualization (equal); Formal Analysis (equal); Investigation (lead); Software (lead), Visualization (lead); Writing – original draft (equal); Writing/Review \& Editing (equal). 
{\bf Sebastian Kempf:} Conceptualization (equal); Formal Analysis (equal); Investigation (supporting); Visualization (supporting); Writing – original draft (equal); Writing/Review \& Editing (equal), Project Administration (lead); Resources (lead); Supervision (lead).

\section*{Data Availability Statement}
The data that support the findings of this study are available from the corresponding author upon reasonable request.

\bibliography{99_bibliography}

\end{document}